\documentclass[10pt,tightenlines,showpacs,letterpaper,aps,prd,twocolumn,nofootinbib,superscriptaddress]{revtex4-2}

\usepackage{amsfonts}
\usepackage{bm}
\usepackage{physics}
\usepackage{graphicx}
\usepackage{xcolor}
\usepackage{multirow}
\usepackage{hyperref}

\begin{document}

\title{Information theoretical perspective on the method of Entanglement Witnesses}

\author{Paulo J. Cavalcanti}
\affiliation{International Centre for Theory of Quantum Technologies, University
of Gdansk, Jana Ba\.zy\'nskiego 1A, Gda\'nsk, 80-309, Poland}

\author{Giovanni Scala}
\email{giovanni.scala@ug.edu.pl}
\affiliation{International Centre for Theory of Quantum Technologies, University
of Gdansk, Jana Ba\.zy\'nskiego 1A, Gda\'nsk, 80-309, Poland}
\affiliation{Dipartimento Interateneo di Fisica, Politecnico di Bari, 70126 Bari, Italy}

\author{Antonio Mandarino}
\affiliation{International Centre for Theory of Quantum Technologies, University
of Gdansk, Jana Ba\.zy\'nskiego 1A, Gda\'nsk, 80-309, Poland}

\author{Cosmo Lupo}
\affiliation{Dipartimento Interateneo di Fisica, Politecnico di Bari, 70126 Bari, Italy}
\affiliation{Dipartimento Interateneo di Fisica, Università di Bari, 70126 Bari, Italy}
\affiliation{INFN, Sezione di Bari, 70126 Bari, Italy}

\begin{abstract}
We frame entanglement detection as a problem of random variable inference to introduce a quantitative method to measure and understand whether 
entanglement witnesses 
lead to an 
efficient procedure for that task. Hence we quantify how many bits of information a family of entanglement witnesses can infer about the entanglement of a given quantum state sample.
The bits are computed 
in terms of the mutual information and we unveil there exists hidden information not \emph{efficiently} processed. We show that there is more information in the expected value of the entanglement witnesses, i.e. $\mathbb{E}[W]=\langle W \rangle_\rho$ than in the sign of $\mathbb{E}[W]$. This suggests that an entanglement witness can provide more information about the entanglement if for our decision boundary we compute a different functional of its expectation value, rather than $\mathrm{sign}\left(\mathbb{E}\right [ W ])$.
\end{abstract}

\maketitle

\section{Introduction}
\label{sec1}
In most of the literature about entanglement detection theory \cite{HHH09,Guehne2009},
the focus is on finding 
separability criteria 
to answer the question of whether a quantum state is entangled or not. 
Efforts along this line of research bring to the development of criteria such as  \emph{positive and partial transposition criterion} PPT \cite{Stoermer1963,Woronowicz1976, Horodecki1996, Horodecki1997}, \emph{realignment criterion} \cite{Zhang2008,chen2003, Sarbickijpa_2020} and 
others \cite{Lupo2008,Sarbicki2020,Aniello2008,Jivulescu2020,Sarbicki2021,Lenny2023,Bera2022}. 

Entanglement detection presents experimental challenges due to the requirement of precise measurements of each individual particle, a task often demanding in practical applications \cite{fid3}. To circumvent this difficulty, \emph{entanglement witnesses} (EW) have been developed \cite{Terhal2000,Terhal2002}, offering a major benefit of providing cost-effective detection without necessitating full tomography of the quantum state.
%
In this manuscript we 
address a 
different question, one that is logically independent of the aforementioned research line discerning whether a state is entangled or not. Instead, 
we posit a query concerning the efficacy of entanglement witnesses:
    what extent of information about the entanglement of a quantum state can be gleaned by computing $\mathrm{sign}[\langle W \rangle_\rho]$, the sign of the expectation value of an entanglement witness?

Subsequently, we seek to determine whether the information garnered from this computation rivals the quantity of information that can be derived from the expectation value itself, that is, how much information is lost by coarse-graining the expectation value to its sign.
 
In order to achieve this, we frame the problem in terms of random variables and evaluate the mutual information between the probabilistic event $\mathcal{E}$ - representing a random quantum state being entangled or not - and $\mathcal{W}$, the expectation value of an entanglement witness on that state, and contrast it with the mutual information between $\mathcal{E}$ and $\mathcal{S} = \text{sign}(\mathcal{W})$, the sign of the expectation value. These two mutual information metrics are then compared to $H(\mathcal{E})$, the entropy of $\mathcal{E}$, which represents the upper limit of knowledge that can be gleaned. So, the output of any protocol that successfully resolves the entanglement detection problem would have mutual information to $\mathcal{E}$ equal to $H(\mathcal{E})$.

Our findings suggest that possessing knowledge of the exact numerical expectation value of the entanglement witness imparts 
much 
more information than merely knowing its sign. However, both methods are still considerably distanced from achieving maximal knowledge. This discrepancy quantifies our distance from resolving the entanglement detection problem using methods based on the entanglement witness, under the assumption that this result does not change qualitatively with the system's dimension $\dim( \mathbb{C}^\mathrm{d_1}\otimes \mathbb{C}^\mathrm{d_2})=\mathrm{d_1}\mathrm{d_2}$.




\section{Entanglement detection as a random variable inference problem}\label{sec2}

In this section, we frame the entanglement witness method as a problem of inference of the value of one random variable through the value of another one.

An entanglement witness is an observable in quantum mechanics which can be used to detect entanglement in a quantum state. Mathematically, it is defined as a hermitian operator $W$, acting on the Hilbert space $\mathcal{H}$, such that its expectation value must be positive $\langle W \rangle_\rho \ge 0$ for all separable quantum states $\rho \in \mathcal{B}(\mathcal{H})$.


This condition can be used to detect entanglement since when the expectation value is negative we know the state must be entangled. Conversely, if the expectation value is positive we have an inconclusive detection. Nonetheless, our goal here is not to detect entanglement. We put the focus on the characterization of the \emph{efficacy of $W$ for entanglement detection} from the information theory perspective. 

In the realm of information theory, Shannon entropy is often employed as a metric to estimate the amount of information required to convey the outcome of a random variable. The entropy of a random variable provides a baseline measure for the average number of bits required to encode the outcomes of said variable \cite{Loomis1949,Chung1986}. For a random variable $Z$ which takes values $z$, this quantity is given by
\begin{equation}
\label{eq:shannonentropy}
    H(Z) = \sum_{z \in Z} p_Z(z) \log_2 p_{Z}(z),
\end{equation}
with $p \log_2 p\equiv 0$ when $p=0$. Should the outcomes of the random variable be deterministic and well-defined, the resultant entropy will be low. Conversely, if the outcomes are spread out over a wide range of equally probable possibilities, the entropy will correspondingly be high, being maximal (given a fixed number of outputs) for the uniform distribution.
Therefore, from an information theory perspective, a significant measure of the mutual dependence of two random variables $X$ and $Y$ is given by the \emph{mutual information}, 
defined in terms of the Shannon entropy Eq.\eqref{eq:shannonentropy} of the two random variables, and reads:
\begin{equation}\label{mutualI}
    I_{XY}(X:Y) = H(X)+H(Y)-H(X,Y),
\end{equation}
where $H(X,Y)$ is the joint entropy computed via $p_{XY}(x,y)$ the joint probability distribution of $X$ and $Y$. 

Mutual information quantifies the amount of information that can in principle be inferred about one random variable from the observation of another random variable. This quantity satisfies $I\geq 0$ and is upper bounded by $\text{min}\{H(X),H(Y)\}$. Our goal is therefore to use it to quantify how much about the entanglement of a state can be inferred from the quantities in the method of entanglement witnesses. 

The random variables whose mutual information values we consider are: $\mathcal{W}$ the expectation value of an entanglement witness on a tossed state, its sign $\mathcal{S}$, and  $\mathcal{E}$ that specifies whether the tossed state is entangled or not. 

More specifically, we randomly generate a state $\rho$ from a sample space\footnote{The set of the density matrices, namely quantum states $\mathcal{B}(\mathcal{H})$ has cardinality infinite, however, in our simulation its subset $\Omega$ is finite.} of quantum states $\Omega\in \mathcal{B}(\mathcal{H})$ and, considering it as a probabilistic event, we define $\mathcal{W}$, $\mathcal{S}$, and $\mathcal{E}$ to take random values $w$, $s$, and $e$ defined as follows:
    \begin{align}
        w &= \mathrm{Tr}(W \rho),\label{eq:X}\\
        s&=\begin{cases}\label{eq:S}
                0 & \text{if }\mathrm{Tr}(W \rho)\ge 0\\
                1 & \text{if }\mathrm{Tr}(W \rho) < 0.
            \end{cases}\\  
        e&=\begin{cases}
                0 & \text{if }\rho\text{ is entangled state}\\
                1 & \text{if }\rho\text{ is separable state}
            \end{cases}
    \end{align}

Notice that, all the events $\mathcal{E},\mathcal{S}$ and $\mathcal{W}$ are probabilistic because $\rho$ is randomly tossed from $\Omega$

Our task, then, is to obtain a quantitative estimation of $I_{\mathcal{W}\mathcal{E}}(\mathcal{W}:\mathcal{E})$, the amount of information about entanglement that can be inferred from $\mathcal{W}$, and of $I_{\mathcal{S}\mathcal{E}}(\mathcal{S}:\mathcal{E})$, which is how much about $\mathcal{E}$ can be inferred from $\text{sign}(\mathcal{W})$. We then proceed to compare them to each other and to $H(\mathcal{E})$, the amount of information required to infer the value of the entanglement random variable with certainty. By varying the entanglement witness $W$, we study the distribution of those mutual information values to reach some conclusion about how informative the set of entanglement witnesses is for the task of entanglement detection.

\begin{figure}
    \centering
    \includegraphics[width=0.9\columnwidth]{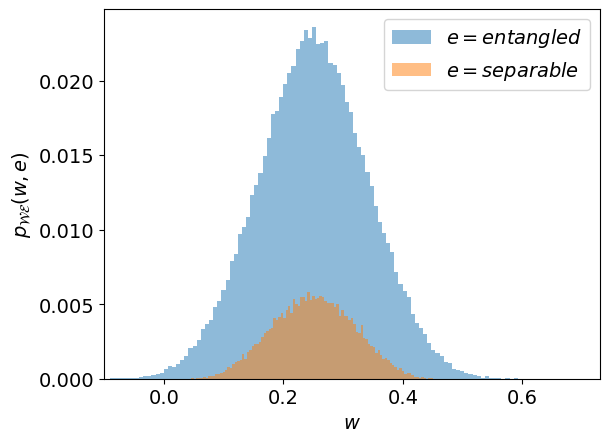}
    \caption{The joint probability $p_\mathcal{WE}(w,e)$ with $e=entangled$ in blue and $e=separable$ in orange for an arbitrary witness $W$ acting on qubit-qubit systems as in Eq. \eqref{eq:Wpartial}. Note that for all separable states $w = \langle W\rangle \ge 0$ and only the entangled states with negative value are detected by this $W$. Note from Eqs. \eqref{eq:X}\eqref{eq:S} that $p_\mathcal{SE}(s,e)$ is a coarse-graning of $p_\mathcal{WE}(w,e)$.}
    \label{fig:distributions}
\end{figure}

To get those estimates, we need to obtain approximations to the joint probability distributions $p_{\mathcal{WE}}$ and $p_{\mathcal{SE}}$ for each of the entanglement witnesses $W$. This can be done only if we select a sample of quantum states for which we know for all $\rho$ whether or not they are entangled. We can achieve that in a particular case by using one of the separability criteria that we mentioned in the introduction, namely, the PPT criterion. By picking the quantum states from the set of qubit-qubit or qubit-qutrit states, the PPT criterion is enough to determine $\mathcal{E}$, and therefore to estimate the probabilities we need. We then numerically generate random states and use the counts of the values of $w$ and $e$ to estimate the probabilities. One example of such $p_{\mathcal{WE}}$ is displayed in Fig. \ref{fig:distributions}. Finally, $p_{\mathcal{SE}}$ is a obtained by coarse-graining $p_{\mathcal{WS}}$ over $w$.

In the following section, we discuss how the random selection of quantum states (necessary for the estimation of $p_\mathcal{WS}$ for a given $W$) and of the operators $W$ (necessary for studying the \textit{set} of such $W$) is performed. 

\section{Sampling quantum states and entanglement witnesses}\label{sec:sampling}

In physics, the symmetries of a studied system are typically known ab initio, despite the dynamics that might be difficult to compute. Since for this particular problem we do not have any specific constraint, we would like to generate random bipartite quantum states \footnote{The code used for the numerical simulations for this work is available at \url{https://github.com/pjcavalcanti/WitnessInformation}. We used tools from the PyTorch and Matplotlib libraries.} from a distribution that is uniform in some sense. We do so by tossing purifications of our states as random rotations of a fiducial pure state, say $\ket{e} = (1,0,...,0)$, where the rotations are chosen from the circular unitary ensemble (CUE), and then tracing out the ancillary system to produce our random bipartite state. The distribution of this ensemble of unitaries is the Haar measure on the unitary group. The Haar measure, in turn, is interesting for us because it is invariant under products with group elements to the left, and therefore can serve as our uniform distribution. Moreover, we toss the matrices $W$ from different sets. Namely, from a family of optimal entanglement witnesses \cite{Chruscinski2014}, from the set of all (up to some normalization) entanglement witnesses, generated with the partial transposition operation, and from a set of random generic hermitian matrices that are not necessarily entanglement witnesses. This provides us with the possibility of making meaningful comparisons.

\subsection{Quantum States}

To sample random states of the bipartite system $\mathbb{C}^{d_\mathrm{A}}\otimes\mathbb{C}^{d_\mathrm{B}}$, we implement the following algorithm \cite{Mezzadri2006,zyczkowski2011}. Let us fix any pure initial state of the system-ancilla combination $\mathbb{C}^{d_A^2 d_B^2} = (\mathbb{C}^{d_\mathrm{A}}\otimes\mathbb{C}^{d_\mathrm{B}}) \otimes (\mathbb{C}^{d_\mathrm{A}}\otimes\mathbb{C}^{d_\mathrm{B}})$, e.g., $\ket{e}=(1,0,\dots,0)$ and then rotate it by $U$, an element  of the circular unitary ensemble that can be obtained as follows. Let $Z\in \mathrm{GL}(d_\mathrm{A}^2d_\mathrm{B}^2,\mathbb{C})$ be a matrix with complex standard normal random variables; feed $Z$ into any QR decomposition routine to get $Z=QR$, where $Q\in \mathcal{U}(d_\mathrm{A}^2d_\mathrm{B}^2)$ and $R$ is invertible and upper-triangular. Here, this decomposition has a \emph{gauge} 
freedom, which, by making the algorithm a multi-valued function, could defeat our goal of defining a uniform distribution. 
Indeed, we could consider $\Lambda=\mathrm{diag}(e^{i\theta_1},\dots,e^{i\theta_N})$ and obtain an equivalent decomposition $Z=Q\Lambda \Lambda^{-1}R$. Therefore, we need to fix a \emph{gauge}, which can be done by choosing $\Lambda=\mathrm{diag}(r_{11}/\lvert r_{11}\rvert,\dots, r_{NN}/\vert r_{NN}\vert)$, 
where $r_{ii}$ are the diagonal elements of $R$. With this, the matrix $U=Q\Lambda$ is uniquely defined and follows a Haar distribution. Then we partial trace the resulting system-ancilla pure state to obtain our random state in $\mathbb{C}^{d_\mathrm{A}}\otimes\mathbb{C}^{d_\mathrm{B}}$,
\begin{equation}
    \rho=\mathrm{Tr}_{\text{ancilla}}\left(U\ket{e}\bra{e}U^\dagger\right).
\end{equation}

\subsection{Observables}
We study the information provided by observables from three families for the qubit-qubit case and two families for the qubit-qutrit case. In this section they are defined, from the higher to the lower expected information according to Fig. \ref{fig:IEX} and \ref{fig:IEW-2x3}.

\subsubsection{Optimal Witnesses}
The first family of witness from which we toss operators $W$ is defined only in the qubit-qubit case and has its members parametrized as follows:

\begin{equation}\label{eq:EW}
    W=a\left(\begin{array}{cc|cc}
    1+\gamma & \cdot & \cdot & \alpha+\beta\\
    \cdot & 1-\gamma & \alpha-\beta & \cdot\\
    \hline \cdot & \alpha-\beta & 1-\gamma & \cdot\\
    \alpha+\beta & \cdot & \cdot & 1+\gamma
    \end{array}\right),\,\lvert\alpha\rvert,\lvert\beta\rvert,\lvert\gamma\rvert\le 1
\end{equation}
and $a>0$ such that $\mathrm{Tr}W=1$ with $\gamma<\sqrt{(\alpha^2+\beta^2/2)}$ \cite{Chruscinski2014}. Here, this condition on $\gamma$ guarantees that all such $W$ have some negative eigenvalue, allowing them, in principle, to detect entanglement for some state. We focus on this particular class of witnesses of Eq. \eqref{eq:EW} because of the characteristic circulant \cite{Chruscinski2014} structure frequently encounter in applications for quantum computing and optics\cite{Knill2001,Johri2020,Mocherla2023}.

\subsubsection{Witnesses from Partial Transposition}
The second family of $W$ we study in both the qubit-qubit and qubit-qutrit cases and has its members constructed as follows:
\begin{equation}\label{eq:Wpartial}
    W=(\bm{1}\otimes T) \left(U\ket{e}\bra{e}U^\dagger\right)
\end{equation}
where $1 \otimes T$ is the partial transposition and $U \in \mathcal{U}(d_A d_B)$ \cite{Chruscinski2014} from the CUE. Since the partial transpose is not completely positive, this can result in some operators $W$ with negative eigenvalues.

\subsubsection{Generic Observables}
To understand how special the entanglement witnesses are for entanglement detection, we repeat our analysis by taking $W$ from a family of Hermitian operators, that is, generic quantum observables, that are not necessarily entanglement witnesses. Hence, by computing the mutual information by the expectation value on a given state there is no theorem that links this value and the entanglement content of the given state.  
Then, the comparison between this mutual information and the one related to the witness allows us to understand, on average, whether a generic observable can extract more information than the witness. 

We generated these generic observables as follows. Given a complex matrix $A$ with standard Gaussian distributed entries, we take its hermitian part and normalize it:
\begin{equation}\label{eq:Wrandom}
    \Tilde{W} = \frac{A + A^\dagger}{|| A + A^\dagger||}
\end{equation}
These operators are naturally defined in both the qubit-qubit and qubit-qutrit cases, allowing them a spectrum with negative support similar to entanglement witnesses.

\section{Results}
\begin{figure}
    \centering
    \includegraphics[width=\columnwidth]{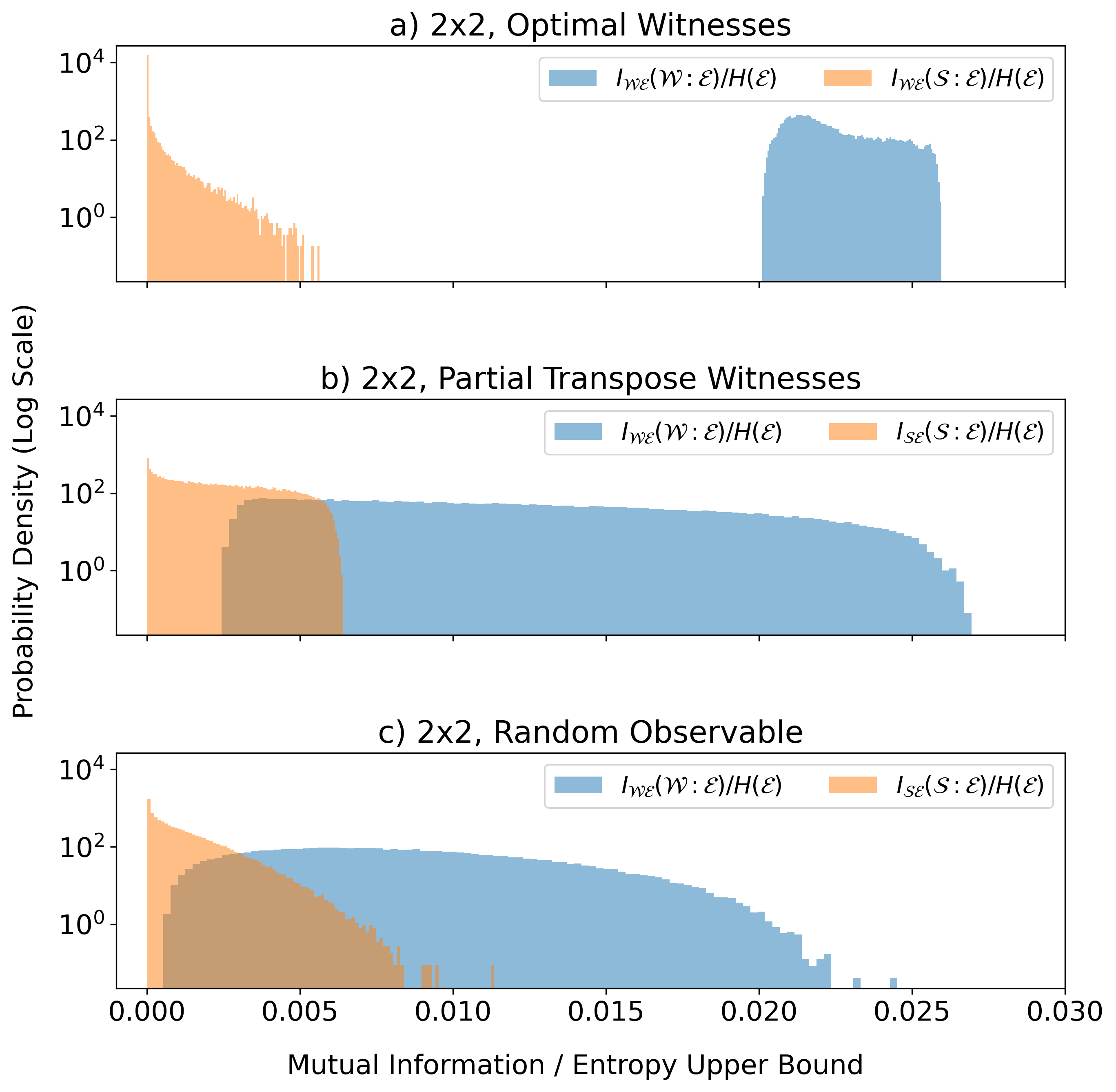}
    \caption{The distribution (log scale) of mutual information values $I_{\mathcal{WE}}(\mathcal{W}:\mathcal{E})$ and $I_{\mathcal{SE}}(\mathcal{S}:\mathcal{E})$ normalized by $H(\mathcal{E})$ in the qubit-qubit case over observables from the three families of interest: a) optimal witnesses of Eq. \eqref{eq:EW}, b) witnesses by partial tracing of Eq. \eqref{eq:Wpartial}, c) random observable of Eq. \eqref{eq:Wrandom}. 
     The probability distribution functions $p_{\mathcal{WE}}(w,e)$ and $p_{\mathcal{E}}(e)$ from which $I_{\mathcal{WE}}(\mathcal{W}:\mathcal{E})$, $I_{\mathcal{SE}}(\mathcal{S}:\mathcal{E})$ and $H(\mathcal{E})$ are computed for each observable are approximated using the counts of the $w,e$ values on a sample of $10^5$ quantum states. The number of histogram bins is arbitrarily chosen to be 100, allocating $10^5$ values of mutual information obtained for each $10^5$ observables $W$ for each family.}
    \label{fig:IEX}
\end{figure}

\begin{figure}[hbt!]
    \centering
    \includegraphics[width=\columnwidth]{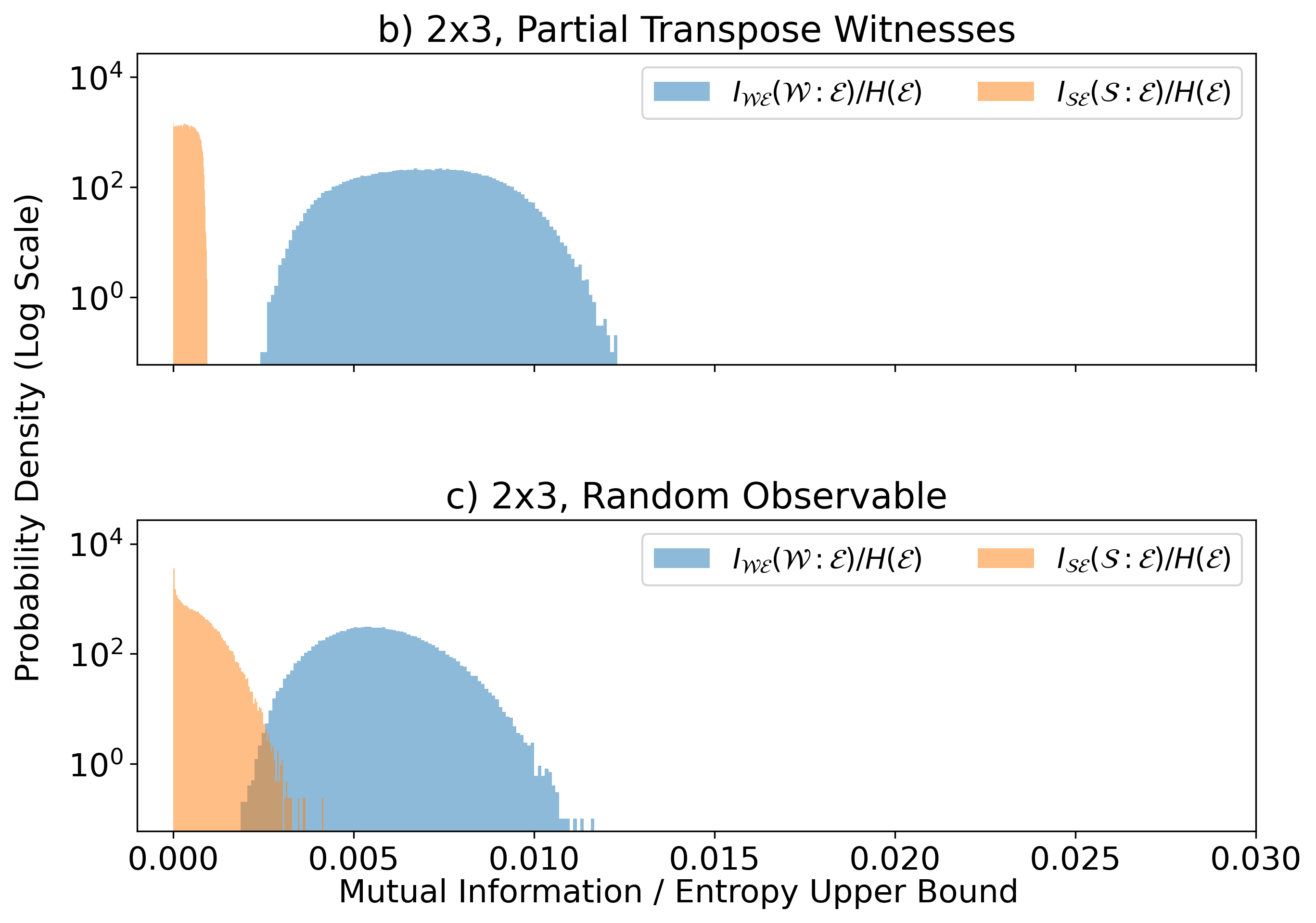}
    \caption{The same quantities (minus optimal witnesses case) as in Fig. \ref{fig:IEX}, repeated for the qubit-qutrit case. The simulations have the same size statistics with $10^5$ observables per family and a sample of $10^5$ states.}
    \label{fig:IEW-2x3}
\end{figure}

We construct the samples of witnesses and observables according to the procedures in the previous section. Given that, we obtain the distributions of mutual information values shown in Figs. \ref{fig:IEX} and \ref{fig:IEW-2x3}, respectively for the qubit-qubit and qubit-qutrit case. 
Specifically, the mutual information probability distributions $I_{\mathcal{WE}}(\mathcal{W}:\mathcal{E})/H(\mathcal{E})$ and $I_{\mathcal{WE}}(\mathcal{W}:\mathcal{E})/H(\mathcal{E})$ reported in the histogram are obtained in the following fashion. 
Once the set of random states is generated with cardinality $10^5$, we numerically estimate the probability distributions $p_{\mathcal{WE}}(w,e)$ and its course-grained $p_{\mathcal{SE}}(s,e)$ for each of the randomly generated witnesses using PPT criterion. 
Later, the probability distributions $p(w,e)$ and $p(s,e)$ referring to the same witness are used to compute a value of the mutual information $I_{\mathcal{WE}}(\mathcal{W}:\mathcal{E})$ and $I_{\mathcal{SE}}(\mathcal{S}:\mathcal{E})$ from Eq. \eqref{mutualI} and \eqref{eq:shannonentropy}. 
Then, repeating the procedure for all the $10^5$ entanglement witnesses we obtain $10^5$ values of $I_{\mathcal{WE}}$ and $I_{\mathcal{SE}}$. 

All of these values are divided by the upper bound for these mutual information values given by $H(\mathcal{E})$, the entropy of the entanglement random variable, giving a qualitative and quantitative comparison of our findings.

Finally, we plot the histograms reporting the distribution of these $10^5$ values of the rescaled mutual information $I_{\mathcal{WE}}/H(\mathcal{E})$ and $I_{\mathcal{SE}}/H(\mathcal{E})$ (with 100 bins, see Fig. \ref{fig:IEX}). 
The distributions on the histograms visualize the ability to infer the entangled or separable character of a typical state given a general witness by looking either at the exact expectation value of the witness computed on the state or just at its sign.

Looking at Fig. \ref{fig:IEX}, we see that the observables from the set of optimal witnesses in Fig.~\ref{fig:IEX}.a on average tend to give a larger fraction of the information than the ones from the set of partial transposition witnesses Fig.~\ref{fig:IEX}.b, which in turn are more informative than the generic observables of Fig.~\ref{fig:IEX}.c. We can also see that in taking the expectation value we gain roughly five times more information than the one available by course graining it with its sign (see also Tabs. I - II). This effect in the coarse-graining seems to be more prominent with the set of optimal witnesses, hinting at a possible trade-off between ease of processing and the amount of information retrieved. We repeat the same analysis for the qubit-qutrit case (minus the set of optimal witnesses, which was defined only for the qubit-qubit systems) and see the same general trends.
It is interesting to notice that, despite the fact that we do not know how to process the information in the expectation value of a generic observable, they are on average almost as informative as the average optimal witness. Quantitatively, from the data used to generate the distribution in Fig. \ref{fig:distributions}, we get that in the qubit-qubit case, $H(\mathcal{E}) \approx 0.80$. We can see the average and standard deviation of the fraction of this information that is given by each class of observables in the following table:

\begin{table}[h]
    \centering
    \caption{2x2 Information Statistics}
    \begin{tabular}{|c|c|c|}
        \hline
         & $I_{\mathcal{WE}}/H(\mathcal{E})$ & $I_{\mathcal{SE}}/H(\mathcal{E})$ \\
        \hline
        \multirow{2}{8em}{Optimal Witnesses} & mean: $0.02$ & mean: $0.00006$ \\
         & std: $0.001$ & std: $0.0003$ \\
        \hline
        \multirow{2}{8em}{Partial Transpose Witnesses} & mean: $0.01 $ & mean: $0.002 $ \\
         & std: $0.007$  & std: 0.002 \\
        \hline
        \multirow{2}{8em}{Random Observables} & mean: $0.008$ & mean: $ 0.001$ \\
         & std: 0.004 & std: 0.001 \\
        \hline
    \end{tabular}
\end{table}

Similarly, for the qubit-qutrit case, we obtain $H(\mathcal{E}) \approx 0.17$ and the information statistics is the following:

\begin{table}[h]
    \centering
    \caption{3x2  Information Statistics}
    \begin{tabular}{|c|c|c|}
        \hline
         & $I_{\mathcal{WE}}/H(\mathcal{E})$ & $I_{\mathcal{SE}}/H(\mathcal{E})$ \\
        \hline
        \multirow{2}{8em}{Partial Transpose Witnesses} & mean: $0.007$ & mean: $0.0004$ \\
         & std: $0.002$ & std: $0.0002$ \\
        \hline
        \multirow{2}{8em}{Random Observables} & mean: $0.006$ & mean: $0.0005$ \\
         & std: $0.001$ & std: $0.0005$ \\
        \hline
    \end{tabular}
\end{table}

We see that in both cases the fraction of the total information extracted is very small, of the order $10^{-2}$, and even less after using the sign of the expectation value as the decision boundary.

\section{Conclusions}
We have addressed how much information can be obtained by knowing the expectation value $\langle W \rangle_{\rho}$ of an entanglement witness $W$, compared to knowing only its sign.
On average, we found one order of magnitude more bits of information with the former method, which indicates that taking the $\mathrm{sign}(\langle W \rangle_{\rho})$ as the decision boundary is a sub-optimal processing of the information available. 
Nevertheless, both methods are still far from being comparable to the entropy $H(\mathcal{E})$.
While our analysis is limited to the simplest qubit-qubit and qubit-qutrit scenarios, it suggests that there might exist a better analysis of the expectation value of the observables (witnesses or not) that can provide more information about the entanglement of a given set of states. The mathematical formulation of this challenge is given by:
\begin{align}\label{eq:problem}
    \arg\max_{\mathbb{F}}I_{\mathbb{F}(\mathrm{Tr}W \rho)\mathcal{E}}(\mathbb{F}(\mathrm{Tr}W \rho):\mathcal{E})
    \, .
\end{align}
Though this issue is earmarked for future research, our current work establishes that the identity function $\mathbb{F}:=\mathrm{id}$ outperforms $\mathbb{F}=\mathrm{sign}$. In geometric terms, using the \emph{sign} gives information about the semi-space of separable states, whereas the \emph{identity function} informs about the projection on the $W$ component parallel to state $\rho$:
\begin{align}
    \langle W\rangle_\rho=\langle W|\rho\rangle_{\mathrm{HS}}=||W||_{\mathrm{HS}}||\rho||_{\mathrm{HS}}\cos\theta \, .
\end{align}
When $\rho$ is pure, $||\rho||_{\mathrm{HS}}=1$. This insight could be crucial for experimental applications, especially for characterizing the tolerance of devices functioning as entanglement sources.


Prompted by our findings, there is potential in refining the processing of the expectation value of entanglement witnesses to harness more information. For instance, a secondary test post observing $\text{sign}(\langle W \rangle_\rho) \geq 0$ could bolster our assurance about the separability of state $\rho$. Notably, this differs from using a second entanglement witness, as it aims to extract more from the same 
observables, 
rather than seeking information from a different witness. 
Going beyond the use of mutual information, future work may explore 
more accurate metrics rooted in hypothesis testing and in the non-asymptotic regime, e.g.~R\'enyi entropies \cite{Tomamichel}.
In a related vein, a complementary study for future research could be addressed to explore the most effective witness or ultrafine witness \cite{ultrafineEW2017} for detecting entanglement for a given state \cite{Shahandeh2014}, as well as nonlocality \cite{Karczewski2022,Ma2018}. This could be achieved by adjusting the parameters of a witness and measuring mutual information 
or other suitable functionals 
as a cost function in a machine learning framework \cite{Hiesmayr2021,MLEW2023,Asif2023,scala2022}.

\section*{Acknowledgments}
We thank A. B. Sainz, G. Sarbicki, and A. Bera for the fruitful scientific discussion.
GS is supported by QuantERA/2/2020, an ERA-Net co-fund in Quantum Technologies, under the eDICT project. PJC and AM acknowledge support by the Foundation for Polish Science (FNP), IRAP project ICTQT, contract no. 2018/MAB/5, co-financed by EU  Smart Growth Operational Programme. CL acknowledges financial support from the PNRR MUR project PE0000023-NQSTI.

\bibliographystyle{apsrev4-2}
\bibliography{MI_EW}


\end{document}